**BioThings Explorer: a query engine for a federated knowledge graph of biomedical APIs**

Jackson Callaghan*,[1], Colleen H. Xu*,[1], Jiwen Xin*,[1], Marco Alvarado Cano[1], Anders Riutta[2], Eric Zhou[1], Rohan Juneja[1], Yao Yao[1], Madhumita Narayan[1], Kristina Hanspers[2], Ayushi Agrawal[2], Alexander R. Pico[2], Chunlei Wu**,[1], Andrew I Su**,[1]

1 Department of Integrative Structural and Computational Biology, The Scripps Research Institute
2 Data Science and Biotechnology, Gladstone Institutes, University of California, San Francisco, San Francisco, CA 94158, USA.

\* - equal contribution
\*\* - corresponding authors

## Abstract

Knowledge graphs are an increasingly common data structure for representing biomedical information. These knowledge graphs can easily represent heterogeneous types of information, and many algorithms and tools exist for querying and analyzing graphs. Biomedical knowledge graphs have been used in a variety of applications, including drug repurposing, identification of drug targets, prediction of drug side effects, and clinical decision support. Typically, knowledge graphs are constructed by centralization and integration of data from multiple disparate sources. Here, we describe BioThings Explorer, an application that can query a virtual, federated knowledge graph derived from the aggregated information in a network of biomedical web services. BioThings Explorer leverages semantically precise annotations of the inputs and outputs for each resource, and automates the chaining of web service calls to execute multi-step graph queries. Because there is no large, centralized knowledge graph to maintain, BioThing Explorer is distributed as a lightweight application that dynamically retrieves information at query time. More information can be found at https://explorer.biothings.io, and code is available at https://github.com/biothings/biothings_explorer.

## Introduction

While downloadable files are the most common way to share biomedical data, application programming interfaces (APIs) are another popular and powerful mechanism for data dissemination. Accessing data through APIs has many complementary advantages relative to downloading local copies of data. APIs typically allow users to query for specific subsets of the data that are of interest. API queries are often highly indexed, leading to efficient data retrieval. Finally, API access allows for easy incorporation of the most up-to-date data into other computational applications or workflows.

While APIs offer many advantages in terms of data accessibility, these advantages do not immediately translate into efficient data integration. APIs generally follow some common architectures and protocols (e.g., representational state transfer (REST), output in JavaScript Object Notation (JSON)), but alignment at this technical level does not guarantee either syntactic or semantic interoperability. For example, APIs can use different identifiers for the same gene, different data structures to represent gene attributes, and different terms to describe the relationships between biomedical entities.

There have been some efforts to define and enforce semantic and syntactic standards to achieve data interoperability. Examples of this approach include the Beacon API from the GA4GH consortium[1] and the DAS specification for sharing annotations of genomic features[2]. These efforts rely on the active participation of API developers in adopting a community standard for their API endpoints.

Here, we explore the use of semantically-precise API annotations as a complementary approach to achieving API interoperability. We divided this work into two components. First, we created an extension of the widely-used OpenAPI annotation standard[3] to semantically describe the APIs' inputs and outputs, and a registry to organize these API annotations. Second, we created an application called BioThings Explorer to consume the API metadata and to execute multi-hop graph queries that span multiple APIs. Together, this architecture allows users to query a large, federated knowledge graph based on an interconnected network of biomedical APIs. This federated design offers a unique approach for creating knowledge graphs that is complementary to the more common strategy of centralization and local data integration.

**A registry of semantically-annotated APIs**

The first step in creating a network of interoperable APIs is to annotate each API in a semantically precise way. We built this API annotation system on the OpenAPI specification[3], the *de facto* standard for documenting API metadata in a human- and machine-readable format. OpenAPI describes basic API metadata (e.g., title, description, version, contact info), as well as key information on the operation of the API endpoints (e.g., server URL, endpoint input parameters, endpoint response schemas).

However, this basic OpenAPI specification does not include key domain-specific information that is necessary to facilitate downstream API interoperability. Therefore, we defined an OpenAPI extension[4] to capture semantically-precise annotations of each API endpoint. These annotations include the semantic types and identifier namespaces of biomedical entities that are both used in querying (inputs) and found in the response (outputs), the JSON path to the output identifier values in the JSON response, and the predicate describing the relationship between the input and output entities (**Supplemental Figure 1**).

We also chose a strategy to map arbitrary JSON data structures to an established biological data model. In this effort, we mapped API output to Biolink Model[5], a community-developed data model that was adopted and extended by the NCATS Translator consortium[6].

To annotate and catalog APIs with our OpenAPI extension, we leveraged the SmartAPI registry[7,8]. The SmartAPI registry currently includes 33 APIs with semantic annotations. This collection of API annotations can be thought of as a "meta-knowledge graph" (meta-KG), where the nodes represent types of biomedical entities (genes, diseases, drugs) and the edges represent APIs that describe relationships between two types of biomedical entities. The SmartAPI meta-KG currently contains 22 nodes and 1041 edges. (The complete meta-KG is shown in **Supplemental Table 1**, and a partial rendering is shown in **Figure 2**).

**API interoperability using BioThings Explorer**

The second step in creating our federated biomedical knowledge graph was to create BioThings Explorer, an engine to autonomously query the SmartAPI meta-KG, query the annotated APIs to retrieve associations between biomedical entities, and integrate those APIs' responses. The input to BioThings Explorer is a query graph, and the syntax for encoding the query graph was defined by the NCATS Translator consortium[6]. The topology of the query graph and the attributes on its nodes and edges define the query (**Figure 1**).

BioThings Explorer executes the query in three distinct phases: query-path planning, query-path execution, and integration and scoring.

Query-path planning. For every edge in a query graph, BioThings Explorer consults the SmartAPI registry for APIs that serve those types of associations (**Figure 1**). For example, in Figure 1, associations between diseases and genes can be found using APIs from the Comparative Toxicogenomics Database[9] and the Biolink API from the Monarch Initiative[5,10], while associations between genes and chemicals can be found using MyChem.Info[11,12]. The sequence of API calls that can satisfy the original query is a "query-path plan".

Query-path execution. In this phase, BioThings Explorer programmatically and autonomously executes each query in each query-path plan based on the semantic annotations for each API identified in the previous phase. BioThings Explorer calls each API, using the SmartAPI annotation to construct calls with the correct syntax and appropriate input identifier, and maps the API responses to the Biolink Model[5]. BioThings Explorer also performs ID-to-object translation, which facilitates the chaining of API calls from one step in the query-path to the next step. This ID translation step is critical when successive APIs in the query-path plan use different identifiers to represent the same biomedical entity (e.g., NCBI Gene ID vs Ensembl Gene ID). The output of this phase is a set of edges for each step of the query-path, which represent the associations between biomedical entities retrieved from the APIs.

<u>Integration and scoring.</u> In this final phase, these sets of edges from the API queries are assembled into result sub-graphs, each of which matches the topology of the query graph. Each result is then scored using a measure of semantic similarity based on the Normalized Google Distance[13].

**Deployment and usage.** The BioThings Explorer knowledge graph is entirely composed from a federated network of APIs. Because there is no local assembly and storage of a large knowledge graph, BioThings Explorer is a very lightweight application that can be easily deployed on almost any standard personal computer. The ability of every user to create a local instance of BioThings Explorer removes centralized bottlenecks associated with large queries and/or heavy usage. The code repository that describes the installation process is at https://github.com/biothings/biothings_explorer.

For users who prefer not to create a local instance of BioThings Explorer, we also maintain a community instance for general use through the NCATS Translator Consortium (https://explorer.biothings.io/).

**Discussion**

Integration of existing data from multiple disparate sources is a key step in assessing the state of current knowledge. There are many existing efforts to create biomedical knowledge graphs by integrating locally-downloaded data and standardizing it using a common data model [14–18]. These efforts result in centralized knowledge graphs of substantial size, often with millions of nodes and tens of millions of edges.

BioThings Explorer offers a unique strategy for data integration, focusing on creating a federated knowledge graph by semantically annotating APIs. Rather than bringing all data into a massive, centralized graph database, this federated design instead allows knowledge to remain behind each resource's API. Data is retrieved at query time by dynamically executing API calls and semantically parsing the results. This architecture functionally separates data dissemination (through API creation) from data modeling and data integration (through semantic annotations).

This approach has several advantages. First, by moving the requirements for interoperability from *implementation in code* to *semantic API annotation*, we significantly lower the barrier to participation in our API ecosystem. Second, by separating these roles into distinct layers, we promote the overall modularity of our system. These components can develop and evolve in parallel, and these two roles can even be undertaken by separate teams (e.g., one team semantically annotates an API that was created by another team). Third, this design facilitates an iterative approach to API annotation. Developers and API annotators can first provide a minimal set of API metadata, which can later be extended based on future needs and use cases.

The federated design of BioThings Explorer also has some notable limitations. First, our OpenAPI extensions in SmartAPI to semantically annotate APIs only work on APIs that follow the REST protocol and provide output in JSON format. Second, because the entire federated KG is never instantiated in a single place, reasoning and scoring methods that rely on having the entire knowledge graph in memory cannot be used with BioThings Explorer.

In sum, we believe that BioThings Explorer is complementary to existing approaches for assembling knowledge graphs, and offers powerful and unique capabilities for both data analysts and tool developers working in biomedical research.

**Availability and Implementation**

BioThings Explorer is implemented as a NodeJS application. The primary repository for the BioThings Explorer project is at https://github.com/biothings/biothings_explorer, which in turn links to and incorporates other repositories as sub-modules. All code is released under the Apache 2.0 open source software license.

# Acknowledgements

Support for this work was provided by the National Center for Advancing Translational Sciences, National Institutes of Health, through the Biomedical Data Translator program, awards OT2TR003427 and OT2TR003445.

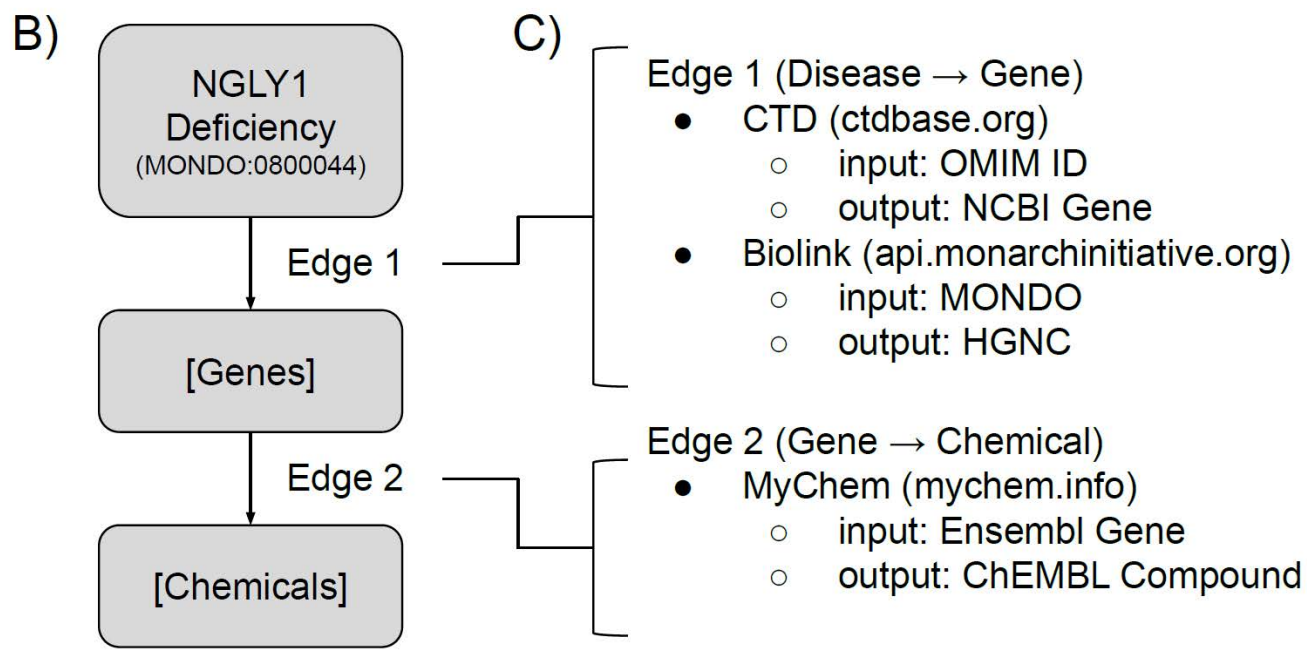

**Figure 1. Deconstruction of a query in BioThings Explorer.** A) An example free-text query that can be answered by BioThings Explorer. B) The graph representation of the same query. The exact syntax of this graph query is specified in the Translator Reasoner API (TRAPI) standard described in Fecho et al.[6] C) The deconstruction of the graph query into multiple API calls by consulting the meta-KG in the SmartAPI registry.

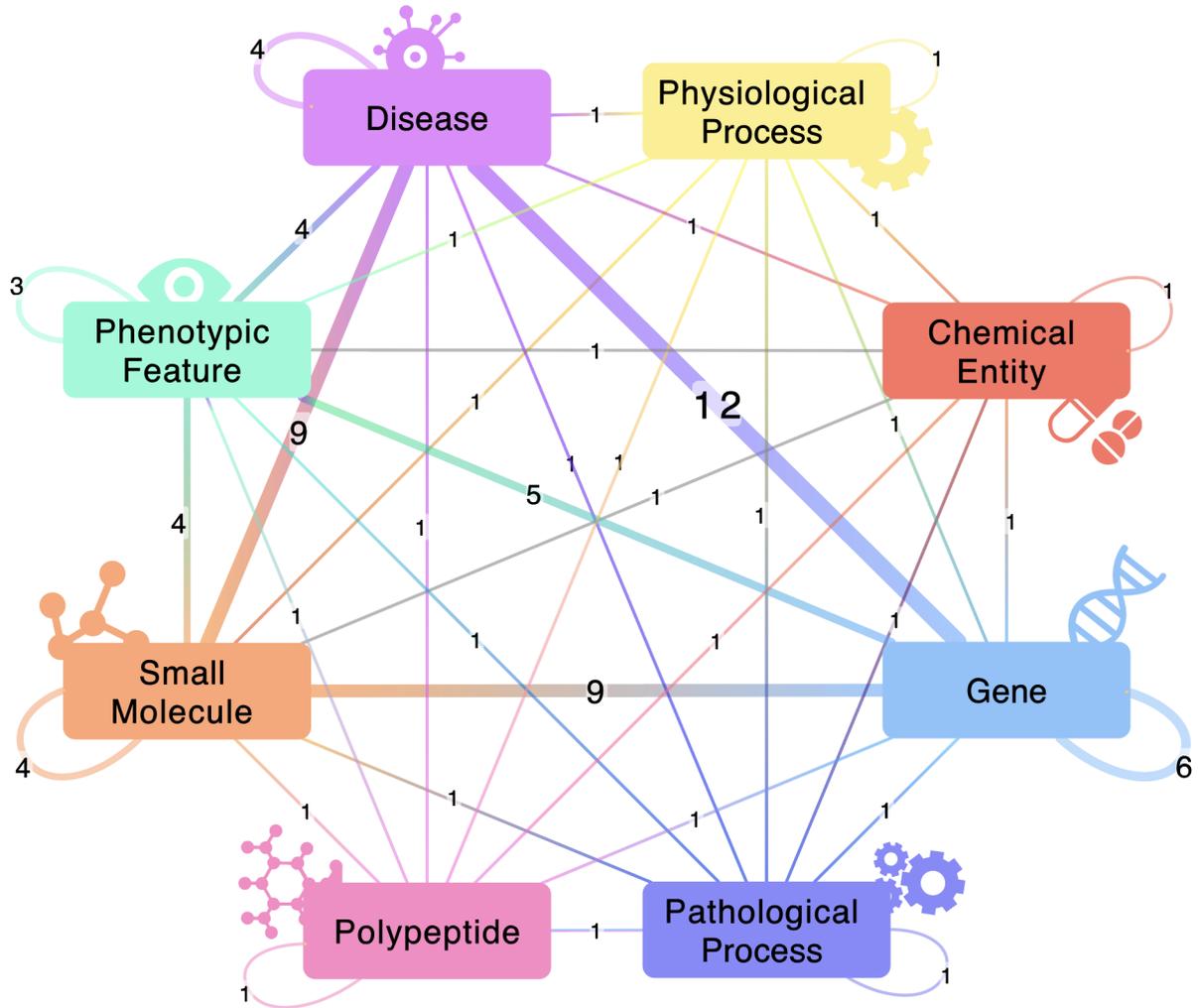

**Figure 2.** A visualization of the meta-knowledge graph for BioThings Explorer. The nodes in this graph are the semantic types of biomedical entities that BioThings Explorer can retrieve associations between (limited to the top 10 most common semantic types). The edges between nodes show what associations between biomedical entities exist in the semantic API network that is accessible through BioThings Explorer. The edge label shows the number of APIs that can retrieve those types of associations, which is also represented by the edge width.

# Supplementary Info

A)

https://www.ncbi.nlm.nih.gov/research/bionlp/litvar/api/v1/entity/litvar/rs121913527%23%23?format=json

B)
```json
{
    "_id": "litvar@rs121913527##",
    "id": "rs121913527##",
    "db": "litvar",
    "years": [ … ], // 15 items
    "diseases": { … }, // 5 items
    "concept": "variant",
    "data": { … }, // 8 items
    "hgvs": "c.146A>T",
    "rsid": "rs121913527",
    "links": [ … ], // 1 item
    "all_hgvs": [ … ], // 18 items
    "hgvs_prot": "p.A146T",
    "weight": 1.996996996996997,
    "pmids_count": 333,
    "first_published_year": 2006,
    "name": "c.146A>T",
    "gene": {
        "id": 3845,
        "name": "KRAS"
    }
}
```

C)
```yaml
x-bte-kgs-operations:
    variant_located_in_gene:
    - supportBatch: false
      useTemplating: true
      inputs:
      - id: DBSNP
        semantic: SequenceVariant
      parameters:
        variantid: "{{ queryInputs | rmPrefix() }}%23%23"   ## no prefix
      outputs:
      - id: NCBIGene
        semantic: Gene
      predicate: is_sequence_variant_of
      source: "infores:dbsnp"
      response_mapping:
        "$ref": "#/components/x-bte-response-mapping/variant_located_in_gene"
x-bte-response-mapping:
    variant_located_in_gene:
      NCBIGene: gene.id  ## no prefix
      "biolink:source_web_page": links.url
```

**Supplemental Figure 1.** Biothings Explorer uses extensions to the OpenAPI specification to semantically annotate APIs. These annotations include the semantic types and identifier namespaces of the biomedical entities that are used in querying (inputs) and found in the response (outputs), the JSON path to the output identifiers in the JSON response, and the predicate describing the relationship between the input and output entities. This figure illustrates these extensions for an API serving content for LitVar. A) The LitVar API can be called using this syntax; the annotation specifies that DBSNP IDs (highlighted in yellow) should be used as the input in queries to this API. B) The API returns a JSON object with a gene related to the variant specified in the query (highlighted in green). C) The `x-kgs-operations` section of the SmartAPI annotation specifies that the data retrieved by the LitVar API relates "SequenceVariants" (identified using DBSNP IDs) to "Genes" (identified using NCBIGene IDs). The `x-bte-response-mapping` portion of the SmartAPI annotation specifies that the NCBIGene IDs of the related genes can be found in the data under the JSON path "gene.id".

**Supplemental Table 1**: https://www.dropbox.com/s/ij09w88zwcuy4uv/supptable.tsv?dl=0